\documentclass[sn-mathphys-num]{sn-jnl}% Math and Physical Sciences Numbered Reference Style 
\usepackage{graphicx}%
\usepackage{multirow}%
\usepackage{amsmath,amssymb,amsfonts}%
\usepackage{amsthm}%
\usepackage{mathrsfs}%
\usepackage[title]{appendix}%
\usepackage{xcolor}%
\usepackage{textcomp}%
\usepackage{manyfoot}%
\usepackage{booktabs}%
\usepackage{algorithm}%
\usepackage{algorithmicx}%
\usepackage{algpseudocode}%
\usepackage{listings}%
\usepackage{natbib}%
\theoremstyle{thmstyleone}%
\theoremstyle{thmstyletwo}%

\theoremstyle{thmstylethree}%

\begin{document}

\title[Article Title]{A variational multi-phase model for elastoplastic materials with microstructure evolution}


\author[1]{\fnm{Sarah} \sur{Dinkelacker-Steinhoff}}\email{sarah.dinkelacker-steinhoff@ruhr-uni-bochum.de}\equalcont{These authors contributed equally to this work.}

\author[1]{\fnm{Klaus} \sur{Hackl}}\email{klaus.hackl@ruhr-uni-bochum.de}
\equalcont{These authors contributed equally to this work.}

\affil[1]{\orgdiv{Institute of Mechanics of Materials}, \orgname{Ruhr-University Bochum}, \orgaddress{\street{Universit\"atsstrasse 150}, \city{Bochum}, \postcode{44801}, \state{North Rhine-Westphalia}, \country{Germany}}}


\abstract{A general model is formulated for elasto-plastic materials undergoing linear kinematic hardening to describe microstructure evolution associated with phase transformations. Using infinitesimal strain theory, the model is based on variational principles for inelastic materials. \\
In our work we combine the so-called dissipation distance, which describes an immediate phase transition in time via an underlying probability matrix. In addition, the volume fractions of the newly emerging phases are represented by Young measures to obtain a time continous microstructure evolution. The model is veryfied employing a two-dimensional benchmark test implemented by the Finite Element Method (FEM). }




\maketitle

\section{Introduction}\label{sec1}

% ähnlich bereits im pamm erschienen
The aim of this work is to provide a detailed insight into a model of microstructure development in relation to phase transformations, which has already been proposed in \cite{DinHa}.
The association of microstructures like for example shear bands with phase transitions, such as density changes, separations or solid-solid transitions (amorphous-amorphous), (amorphous-cristalline) occur in different materials and scales. These include soils,  amorphous metals or  vitreous silica, the coupled appearance of which is not yet fully understood. 
Within the framework of continuum mechanics, similarities between the various natural and construction materials can be identified on a macroscopic level. Even if the intrinsically acting forces are different, the lack of long-range order is mentioned as an explanation for the supposedly similar plastic deformation behavior. In this context, the similarity of the yield behavior is used, which can be modeled by ellipsoidal yield surfaces in metallic glasses and amorphous oxides. Therefore, Mohr-Coulomb or Drucker-Prager-like models can be found in literature for the application of numerical analyses \cite{Anand,Molnar,Bruns}.  \\
Even though there are many examples of the respective materials providing phase transformations coupled with microstructure formations, a few will be mentioned below.
For instance, molecular dynamics (MD) simulations and high-pressure experiments have shown over the years that the density of fused silica changes during compression.
In the pressure range from 0 to 140 GPa, the coordination number (CN) of Si chances from 4-fold to 6-fold \cite{Petitgirard}. Apart from the transformation in the short-range order, restructuring can also be observed in the medium range. The ring size distribution was described in terms of elastic and plastic compression in D\'avila et al. \cite{Davila}, whereby the distribution is subject to continuous change during plastic deformation. A ring is here considered to be a group of Si-O with the shortest bond distance.\\
Additionally, Schill et al. \cite{Schill1} interpreted the anomalous behaviour of the elastic moduli and an anomalous critical state line as an indication of microstructure formation. 
There are also measurements and simulations of shearbands from bulk metallic glasses (BMGs) \cite{Greer,He}. For instance, Mironchuk et al. \cite{Mironchuk} observed phase separations up to transformations in the range from (10-20) nm in Al$_{87}$Ni$_8$La$_5$ alloy during a compression test. Likewise, a coarsening of the microstructure at shear zones was observed in various BMGs. In order to be able to investigate their development and the effects more precisely, for example synchrotron XRD measurements were carried out on metallic glasses that had previously been provided with nanospheres. See, \cite{He} for more details. \\
In contrast to this materials, shear bands in granular substances are optically visible and a development of volume change in the affected regions during shear stress is well-known \cite{Tafili, Wolf}. The effect of decreasing density and thus increasing volume, so-called dilatancy, is due to local plastic flow.\\
The variational framework used in this work has already been applied under different aspects \cite{Behr,Kochmann1,Hackl1}. Approaches empolying a stochastic probability matrix can be found in \cite{Gov, MWolff}. Here, we use a combination of both approaches together with the concepts of dissipation distance and generalised Young measures to better understand microstructure evolution which are coupled with phase transformations.
\section{Variational model}
%non-conservativ System

From the perspective of continuum mechanics, a general non-conservative system can dissipate energy during deformation. This change affects the evolution of the internal material variables. Therefore, two potentials that depend on two different types of state variables are often used to describe the thermodynamic evolution. In the following case, the Helmholz free energy density
$\Psi(\boldsymbol{\varepsilon},\boldsymbol{z})$ and the dissipation potential $\Delta(\boldsymbol{z}, \dot{\boldsymbol{ z}})$ depend on the external controllable total strain $\boldsymbol{\varepsilon}$ and the internal state variables $\boldsymbol{z}$.\\
To examine the conditions of equilibrium state it is possible to use the behavior known from the second law of thermodynamics that a system wants to maximize its entropy. This correlates with the minimum of free energy. In addition to this principle, thermodynamic stability leads to the inclusion  
\begin{align}\label{eq:eq1}
\boldsymbol{0} \in \frac{\partial \Psi}{\partial \boldsymbol{ z}}
+ \frac{\partial \Delta}{\partial  \dot{\boldsymbol{ z}}} .
\end{align}
A similar description can be traced back to Biot, whereby conditions of convex analysis are considered here.
Equation \eqref{eq:eq1} can be reformulated minimizing the Lagragian $\mathcal{L} = \dot \Psi+ \Delta $ with respect to the time derivative of the internal variable $\boldsymbol{z}$. This is known as principle of the minimum of the dissipation potential for inelastic materials and has already been discussed in detail in the literature \cite{Hackl1, Carstensen, Ortiz_Repetto}. 
Our variational model is based on these minimum principles. 
In the following, a classical variational description of a material with a single phase will be discussed first, followed by a more detailed discussion of a general multi-phase approach.

\subsection{Classic description for a single phase}
\label{chap:chap1.2}
% standard Helmholz free energy
At first, the two potentials introduced above have to be defined. If we asume an elasto-plastic material with linear kinematic hardening response, then a
standard Helmholz energy function $\Psi(\boldsymbol{\varepsilon},\boldsymbol{\varepsilon}^\mathrm{p})$ can be build of an elastic energy term and a term which describes the linear kinematic hardening
\begin{align}\label{eq:eq2}
 \Psi(\boldsymbol{\varepsilon},\boldsymbol{\varepsilon}^\mathrm{p}) = \frac{1}{2} (\boldsymbol{\varepsilon} -\boldsymbol{\varepsilon}^\mathrm{p} )\colon \mathbb{C} \colon (\boldsymbol{\varepsilon} -\boldsymbol{\varepsilon}^\mathrm{p}) + \frac{b}{2} \lVert{\boldsymbol{\varepsilon}^\mathrm{p}}\rVert ^{2}.
\end{align}
In this formulation linear strain theory with the decompostion of the total strain tensor $\boldsymbol{\varepsilon}$ into an elastic and a plastic part is used
\begin{align}\label{eq:eq3}
\boldsymbol{\varepsilon}= \frac{1}{2} ( \nabla \boldsymbol{\mathrm{u}} + \nabla \boldsymbol{\mathrm{u}}^{\mathrm{T}}) = \boldsymbol{\varepsilon}^\mathrm{e} + \boldsymbol{\varepsilon}^\mathrm{p}. 
\end{align}
Moreover, the free energy contains the fourth order stiffness tensor $\mathbb{C}$. It should be noted that in the standard case the hardening term only consists of the plastic strain tensor and the parameter $b$ is a constant with $b>0$
Moreover we asume plastic incompressibility, which leads to the constraint $\mathrm{tr}(\boldsymbol{\varepsilon}^\mathrm{p}) = 0$ in the material model.\\
%evolution, dissipation
In a simple case of dry friction, the dissipation potential $\Delta(\dot { \boldsymbol{\varepsilon}}^\mathrm{p})$
depends on the rate of the plastic strain tensor and $r>0 $ represents a dissipation constant
\begin{align}\label{eq:eq4}
\Delta = r \lVert \dot{\boldsymbol{ \varepsilon}}^\mathrm{p} \rVert, 
\end{align} 
see, e.g. \cite{HF}.\\
Moreover, this process is rate-independent because the dissipation potential is homogeneous of first order in the argument. Hackl and Fischer \cite{HF} have shown that in this case instead of the often described maximum principle, see for instands \cite{Miehe}, the principle of the minimum of the dissipation potential can be applied giving the formula
\begin{align}
\dot{\boldsymbol{ \varepsilon}}^\mathrm{p}= \arg \min\limits_{\{ \dot{\boldsymbol{ \varepsilon}}_i^\mathrm{p}\}} \left\{ \mathcal{L^*}(\boldsymbol{\varepsilon},\boldsymbol{ \varepsilon}^\mathrm{p} \dot{\boldsymbol{  \varepsilon}}^\mathrm{p})\right\}.
\end{align}
The Lagrangian 
\begin{align}
\mathcal{L^*}(\boldsymbol{\varepsilon},\boldsymbol{ \varepsilon}^\mathrm{p}, \dot{\boldsymbol{  \varepsilon}}^\mathrm{p}) = \dot \Psi+ \Delta
= \dot{\boldsymbol{\varepsilon}} \colon \mathbb{C} \colon (\boldsymbol{\varepsilon} -\boldsymbol{\varepsilon}^\mathrm{p})-\dot{\boldsymbol{  \varepsilon}}^\mathrm{p} \colon \mathbb{C} \colon (\mathrm{dev}(\boldsymbol{\varepsilon}) -\boldsymbol{\varepsilon}^\mathrm{p}) + b~\dot{ \boldsymbol{  \varepsilon}}^\mathrm{p} \colon \boldsymbol{\varepsilon}^\mathrm{p} + r \lVert \dot{\boldsymbol{ \varepsilon}}^\mathrm{p} \rVert
\end{align}  
is the summation of  the total derivative of the free energy density $\dot \Psi$ and the forementioned dissipation potential in Eq.\eqref{eq:eq4}.
The variational approach results with $0 \in \partial \mathcal{L} $ in the differential inclusion   
\begin{align}\label{eq:eq5}
\mathrm{dev} \boldsymbol{\sigma} - b~ \boldsymbol{\varepsilon}^\mathrm{p} \in r ~ \mathrm{sign} (\dot{\boldsymbol{ \varepsilon}}^\mathrm{p} ),
\end{align}
which can solved explicitly.
Introducing the yield function
\begin{align}
	\phi = \lVert \mathrm{dev} \boldsymbol{\sigma} - b \, \boldsymbol{\varepsilon}^\mathrm{p} \rVert -r,
\end{align}
Eq.~\eqref{eq:eq5} is equivalent to the flow rule
 \begin{align}
\dot{\boldsymbol{\varepsilon}}^\mathrm{p}= \mu ( \mathrm{dev} \boldsymbol{\sigma} - b \, \boldsymbol{\varepsilon}^\mathrm{p}),
\end{align}
complimented by the Kuhn-Tucker conditions 
\begin{align}
	\mu\geq 0,~ \phi \leq 0,~ \mu\,\theta= 0.
\end{align}
The flow rule describes the evolution of the internal variable. Only the deviatoric stress remains after variation, and $\mu $ denotes a Lagrange multiplier, the so-called consistency parameter. 

\subsection{General multi-phase approach }

If a material is capable of forming several phases during the deformation process, the mathematical description depends on various factors. If the material is crystalline then there are for example lattice effects to consider, like dislocations, or polymorphic transitions. \\
We restrict ourselves to a homogeneous material which is subject to an isothermal change of state. During deformation, isotropic phase changes can occur. 

\paragraph{Generalised free energy density}

In general, phase mixing resembles a stochastic process that describes the kinetics of transformation. In our case, an initial phase may evolve into other phases.
We assume a material with $k$ possible phases and volume fractions $\lambda_i$, with $i \in \{1, ... ,k\}$.
Moreover, for a closed system, mass conservation can be formulated as
\begin{align}
\sum_i^k \lambda_i =1, \hspace{0.5cm} 0\leq \lambda_i \leq1.\label{eq:eq_mass}
\end{align}
%During the time development constant volume fractions $\lambda_i$ occur instantaneously and one step during phase changing can be physically seen as a state.\\
The transition from a state $i$ to a diffrent state $j$ can be described by its rate $g_{ij}$. This gives rise to a transition matrix $\boldsymbol{g} \in \mathbb{R}^{k \times k}$, which governs the stochatic progress of the phase development:
\begin{equation}
\boldsymbol{g} = \begin{bmatrix}
   g_{11} & g_{12} & \cdots & g_{1k} \\
   g_{21} & g_{22} & \cdots & g_{2k} \\
   \vdots  & \vdots  & \ddots & \vdots  \\
   g_{k1} & g_{k2} & \cdots & g_{kk} 
 \end{bmatrix}\label{eq:mat}
\end{equation}
%These states, $i$ and $j$, are in turn filled with all possible phases. Therefore we write in short $ i,j \in \{ 1, ...,k\} \subset \mathbb{N}$ .\\
We set $g_{ii} = 0$, because there are no transitions between a certain phase and itself. Moreover, we assume $g_{ij}\geq0$, i.e., transition from phase $i$ to phase $j$ is discribed by $g_{ij}$, and back-transition from phase $j$ to phase $i$ is discribed by $g_{ji}$ only. Note that for now, transion in both directions may occur simultaniously. Later on, this will be modified.\\
The rate of phase change is given by
\begin{align}
	\dot \lambda_i  = \sum_j^k \left( g_{ji} - g_{ij}. \right)\label{eq:eq6}
\end{align}
a general connection to the rate of each phase is given. For the sake of simplicity the focus is set on the development from $i \rightarrow j$. 
The sum in Eq. \eqref{eq:eq_mass} is guaranteed to be zero this way.

In the following, the total Helmholtz free energy density is defined as the average of the energy introduced in Eq. \eqref{eq:eq2} of all individual phases. This gives,
\begin{align}
\Psi_i &= \frac{1}{2} (\boldsymbol{\varepsilon}_i -\boldsymbol{\varepsilon}^\mathrm{p}_i)\colon \mathbb{C}_i  \colon (\boldsymbol{\varepsilon}_i -\boldsymbol{\varepsilon}^\mathrm{p}_i) + \frac{b}{2} \lVert{\boldsymbol{\varepsilon}^\mathrm{p}_i}\rVert ^{2} + c_i, \nonumber 
\\ 
\Psi_\mathrm{{tot}}&= \sum_{i}^{k} \lambda_i \Psi_i.\label{Eq:Eq.8}
\end{align}
%At this point it is important to note that although the states are not physically the same, the indices $i$ and $j$ are interchangeable to find the energy in the $j_{th}$ state. Therefore $\Psi_{j}$ has also be defined to find the driving forces later on.
\paragraph{ Dissipation potentials}
In order to model inealastic evolution, we introduce three types of dissipation potantials resposible for the various processes as
\begin{align} \label{eq:9}
			\Delta_{\mathrm{plast}}&= \sum_i 
			\lambda_i r_i \rVert  \dot{\boldsymbol{ \varepsilon}}^\mathrm{p}_i \lVert
			\\\label{eq:11}
			\Delta_{\mathrm{trans}}&= \sum_{i,j} g_{ij} D_i\left( \boldsymbol{ \varepsilon}_i^\mathrm{p}, \boldsymbol{ \varepsilon}_j^\mathrm{p}\right)
			\\ \label{eq:10}
			\Delta_{\mathrm{reg}}&= \frac{1}{2 \eta_1} \sum_i \lambda_i \rVert \dot{\boldsymbol{  \varepsilon}}^\mathrm{p}_i \lVert^2 + \frac{1}{2 \eta_2} \sum_{i,j} g_{ij}^2
\end{align}
Here, $\Delta_{\mathrm{plast}}$ describes plastic deformation in the individual phases according to Eq.~\eqref{eq:eq4}. Note, that the different phases may possess different yield limits. 

Furthermore, $\Delta_{\mathrm{trans}}$ governs the transformation between the individual phases. We assume the transition between the states to be immidiate. This can be modeled employing a so-called dissipation distance $D (\boldsymbol{z_0},\boldsymbol{z_1})$, which is related to the dissipation potential $\Delta(\boldsymbol{z_0})$ of the corresponding process via 
\begin{align}
	\label{eq:eq09a}
D(\boldsymbol{z_0},\boldsymbol{z_1}) =\inf\limits_{\boldsymbol{z}} \{ \int_{t_0}^{t_1} \Delta(\boldsymbol{z_0},\dot{\boldsymbol{ z}})~ \mathrm{d}t \vert~ \boldsymbol{z}(t_0) = \boldsymbol{z_0}, \boldsymbol{z}(t_1)=\boldsymbol{z}_1
   \},
\end{align}
see \cite{Mielke04,Kochmann1,JUNKER2025} for further information.\\
In this work we assume a dissipation potential of the form
$
\Delta_i(\boldsymbol{\varepsilon}^\mathrm{p},\dot{\boldsymbol{  \varepsilon}}^\mathrm{p})= ~r_i~ \rVert \dot{\boldsymbol{  \varepsilon}}^\mathrm{p} \lVert.
$
The dissipation distance related to state $i$ can be obtained by path minimization of the dissipation potential, as indicated in Eq.~\eqref{eq:eq09a}. This yields
\begin{align}
D_i\left( \boldsymbol{ \varepsilon}_{\mathrm{start}}^\mathrm{p}, \boldsymbol{ \varepsilon}_{\mathrm{end}}^\mathrm{p}\right)&=\inf\limits_{\boldsymbol{\varepsilon}^\mathrm{p}} \{ \int_{t_0}^{t_1} \Delta_i(\boldsymbol{\varepsilon}^\mathrm{p},\dot{\boldsymbol{  \varepsilon}}^\mathrm{p})~ \mathrm{d}t \vert~  \boldsymbol{\varepsilon}^\mathrm{p}(t_0) = \boldsymbol{\varepsilon}^\mathrm{p}_{\mathrm{start}}, \boldsymbol{\varepsilon}^\mathrm{p}(t_1)=\boldsymbol{\varepsilon}^\mathrm{p}_{\mathrm{end}}
   \} \nonumber \\
&= r_i \lVert \boldsymbol{ \varepsilon}_{\mathrm{end}}^\mathrm{p} - \boldsymbol{\varepsilon}_{\mathrm{start}}^\mathrm{p} \rVert\label{eq:eq13}.
\end{align}
%whereas in our case $\boldsymbol{  \varepsilon}^\mathrm{p}_{\mathrm{start}} = \boldsymbol{  \varepsilon}^\mathrm{p}_i $ and $\boldsymbol{  \varepsilon}^\mathrm{p}_{\mathrm{end}} = \boldsymbol{  \varepsilon}^\mathrm{p}_j$.

Finally, we add a viscous regularization given by $\Delta_{\mathrm{reg}}$. This term is not strictly required, but will simplify mathematical treatment later on and also renders the computational simulation more robust.

%In addition to the homogenous case with order of one in the rate $\boldsymbol{ \dot \varepsilon}^\mathrm{p}$, a viscoplastic process as regularization term Eq. \eqref{eq:10} is added. It has the important characterics that the derived evolution equations are getting explicit and the quadratic terms results in a rate dependend behaviour, with the viscosity parameters $\frac{1}{\eta_s}$, $s \in \{ \mathrm{1,2} \}$. In this case it not restrict the application of the minimum principle \cite{HF}.\\

\subsection{Relaxed energy}
%The previous concept leads to the condensed energy potential often mentioned in the literature \cite{Mielke04, Kochmann1}.  
In order to describe a continous evolution of the internal variables for an existing microstructure, a relaxation concept via Young measures can be used. This has been discussed in \cite{Kochmann1} for gradient Young measures.

Let us assume a general Young measure describing the distribution of an internal variable $\boldsymbol{z}$ of the form
\begin{align}
 f(\boldsymbol{z})= \sum\limits_{i=1}^k \lambda_i ~\delta( \boldsymbol{z}-\boldsymbol{z}_i),
\end{align}
i.e., as a discrete distribution with respect to fixed volume fractions.

In this work, we restrict ourselves to macroscopic information. Then a relaxed, i.e. effective, free energy potential can be determined via
\begin{align}
\Psi_{\mathrm{rel}}(\boldsymbol{\varepsilon},\boldsymbol{\varepsilon}_{\mathrm{eff}}^\mathrm{p})= \min\{\Psi_{\mathrm{tot}}~ \vert~ \boldsymbol{\varepsilon}_i ,~ \sum_i \lambda_i \boldsymbol{\varepsilon}_i~ = ~ \boldsymbol{\varepsilon}\}.
\end{align}
Minimizing the total Helmholz energy density in Eq.~\eqref{Eq:Eq.8} with respect to the averaged strain components over each volume fraction $\boldsymbol{\varepsilon}_i$, and considering the constraint $\sum_{i=1}^k \lambda_i \boldsymbol{\varepsilon}_i = \boldsymbol{\varepsilon}$
the effectiv tensors 
\begin{align}
 \mathbb{C}_{\mathrm{eff}} &= \left(\sum_i \lambda_i \mathbb{C}_i^{-1}\right)^{-1}, ~~ \boldsymbol{\varepsilon}_{\mathrm{eff}}^\mathrm{p} = \sum_i \lambda_i \boldsymbol{\varepsilon}_i^\mathrm{p}
\end{align}
 are obtained. The effective stiffness tensor $\mathbb{C}_{\mathrm{eff}} $ relates to a homogenous \textit{Reu\ss} lower bound formulation, see e.g. \cite{Behr}. This includes the stresses being spatially constant at the material point level. The relaxed energy is obtained as
\begin{align}
\Psi_{\mathrm{rel}}(\boldsymbol{\varepsilon},\boldsymbol{\varepsilon}_i^\mathrm{p},\lambda_i)= \frac{1}{2} (\boldsymbol{\varepsilon} -\boldsymbol{\varepsilon}_{\mathrm{eff}}^\mathrm{p} )\colon \mathbb{C}_{\mathrm{eff}}  \colon (\boldsymbol{\varepsilon} -\boldsymbol{\varepsilon}_{\mathrm{eff}}^\mathrm{p}) + \frac{1}{2} \sum\limits_i^k \lambda_i b_i  \lVert{\boldsymbol{\varepsilon}_i^\mathrm{p}}\rVert ^{2} + \sum\limits_i^k \lambda_i c_i.
\end{align} 
And the related stress is given via the equation 

\begin{align}
\boldsymbol{\sigma}= \frac{\partial \Psi_{\mathrm{rel}}} {\partial \boldsymbol{\varepsilon}} = \mathbb{C}_{\mathrm{eff}}  \colon (\boldsymbol{\varepsilon} -\boldsymbol{\varepsilon}_{\mathrm{eff}}^\mathrm{p}).
\label{eq101}
\end{align}

\subsection{Evolution equations }
We follow the variational formulation established in Section \ref{chap:chap1.2}, where the Lagrange functional has to be minimized with respect to the kinetic variables $\dot{\boldsymbol{  \varepsilon}}_i^\mathrm{p}$ and $g_{ij}$ as
\begin{align}
\min\limits_{\{\dot{\boldsymbol{  \varepsilon}}_i^\mathrm{p}\}, \{g_{ij}\ge 0\}} \left(  \dot \Psi_{\mathrm{rel}} + \Delta_{\mathrm{reg}} + \Delta_{\mathrm{trans}} + \Delta_{\mathrm{plast}}\right).
\end{align}
Employing Eqs. ~\eqref{eq:eq6} and \eqref{eq101} the time derivative of the relaxed energy can be written as
\begin{align}
\dot \Psi_{\mathrm{rel}}= \frac{\mathrm{d}}{\mathrm{d}t}\sum\limits_i^k \Psi_{\mathrm{rel}} (\boldsymbol{\varepsilon},\boldsymbol{\varepsilon}_i^\mathrm{p},\lambda_i)=\boldsymbol{\sigma} \colon \dot{\boldsymbol{ \varepsilon}} + \sum\limits_i^k \left[ \frac{\partial \Psi_{\mathrm{rel}}}{ \partial \boldsymbol{\varepsilon}_i^\mathrm{p}} \colon  \dot{\boldsymbol{ \varepsilon}}_i^\mathrm{p} +\frac{\partial \Psi_{\mathrm{rel}}}{ \partial \lambda_i} \sum_j^k \left( g_{ji} - g_{ij}\right) \right]. 
\label{eq102}
\end{align}
The thermodynamically conjungated driving forces for phase change
 $\partial \Psi_{\mathrm{rel}} / \partial \lambda_i$ and $\partial \Psi_{\mathrm{rel}} / \partial \boldsymbol{\varepsilon}_i^\mathrm{p}$ can be determined.
The corresponding Lagrangian $\mathcal{L}$ reads, neglecting the leading term in Eq.~\eqref{eq102}, which is constant with respect to variation,
 \begin{multline}
\mathcal{L}= \sum\limits_{i} \left[ -\lambda_i \dot{\boldsymbol{ \varepsilon}}_i^\mathrm{p} \colon \mathbb{C}_{\mathrm{eff}}  \colon (\boldsymbol{\varepsilon} -\boldsymbol{\varepsilon}_{\mathrm{eff}}^\mathrm{p}) + \dfrac{\partial \Psi_{\mathrm{rel}}}{ \partial\lambda_i} \sum_j \left( g_{ji} - g_{ij}\right) + \lambda_i b_i \boldsymbol{\varepsilon}_i^\mathrm{p} \colon \dot{\boldsymbol{ \varepsilon}}_i^\mathrm{p}  \right] \\
+ \sum\limits_i \lambda_i r_i \rVert \dot{\boldsymbol{  \varepsilon}}^\mathrm{p}_i \lVert +
 \sum\limits_j g_{ij} r_i \lVert{\boldsymbol{\varepsilon}_j^\mathrm{p}} - {\boldsymbol{\varepsilon}_i^\mathrm{p}}\rVert 
      + \dfrac{1}{2 \eta_1}\sum\limits_i \lambda_i \lVert \boldsymbol{\varepsilon}_i^\mathrm{p} \rVert^2 
      + \dfrac{1}{2 \eta_2} \sum\limits_{i,j} g_{ij} ^2.
\end{multline}
Stationarity requires 
\begin{align}
 \mathbf{0} \in \frac{\partial \mathcal{L}}{\partial \dot {\boldsymbol{  \varepsilon}}_i^\mathrm{p}},~
 0\in \frac{\partial \mathcal{L}}{\partial g_{ij}}
 \end{align}
and leads first to the differential inclusions  
\begin{align}
 \mathbb{C}_{\mathrm{eff}}  \colon (\mathrm{dev} (\boldsymbol{\varepsilon}) -\boldsymbol{\varepsilon}_{\mathrm{eff}}^\mathrm{p}) - b_i \boldsymbol{\varepsilon}_i^\mathrm{p}-\frac{1}{\eta_1}\dot{\boldsymbol{\varepsilon}}_i^\mathrm{p}  \in r_i \, \mathrm{sign}( \dot{\boldsymbol{\varepsilon}}_i^\mathrm{p})
 \end{align}
and 
 \begin{align}
 \dfrac{\partial \Psi_{\mathrm{rel}}}{ \partial\lambda_i}-\dfrac{\partial \Psi_{\mathrm{rel}}}{ \partial\lambda_j} - \frac{1}{\eta_2} g_{ij}
    \in  r_i \lVert \boldsymbol{ \varepsilon}_j^\mathrm{p}-\boldsymbol{\varepsilon}_i^\mathrm{p} \rVert,  
 \end{align}
which can be solved for $\dot{\boldsymbol{ \varepsilon}}_i^\mathrm{p}$ and $g_{ij}$, giving the evolution equations
\begin{align}
\dot{\boldsymbol{ \varepsilon}}_i^\mathrm{p} &= \eta_1 \left( \Vert \mathbb{C}_{\mathrm{eff}}  \colon (\mathrm{dev} (\boldsymbol{\varepsilon}) -\boldsymbol{\varepsilon}_{\mathrm{eff}}^\mathrm{p}) - b_i \boldsymbol{\varepsilon}_i^\mathrm{p}\rVert  -r_i \right)_+ \, \mathrm{sign} \left( \mathbb{C}_{\mathrm{eff}}  \colon (\mathrm{dev} (\boldsymbol{\varepsilon}) -\boldsymbol{\varepsilon}_{\mathrm{eff}}^\mathrm{p})\right), \label{eq:eq19a} \\
g_{ij}&= \eta_2 
    \left( \dfrac{\partial \Psi_{\mathrm{rel}}}{ \partial\lambda_i}-\dfrac{\partial \Psi_{\mathrm{rel}}}{ \partial\lambda_j}
    - r_i \lVert \boldsymbol{ \varepsilon}_j^\mathrm{p}-\boldsymbol{\varepsilon}_i^\mathrm{p} \rVert \right)_+ . \label{eq:eq19}
\end{align}
Here, $(a)_+=\mathrm{max}(a,0)$ denotes the positive value of the argument. Note that Eq.~\eqref{eq:eq19} ensures that always $g_{ij}=0$ or $g_{ji}=0$. Hence, there are never phase transions in both directions possible, cancelling each other. Equations~\eqref{eq:eq19a} and \eqref{eq:eq19} give in a natural way rise to yield functions
\begin{align}
	\phi_i = \Vert \mathbb{C}_{\mathrm{eff}}  \colon (\mathrm{dev} (\boldsymbol{\varepsilon}) -\boldsymbol{\varepsilon}_{\mathrm{eff}}^\mathrm{p}) - b_i \boldsymbol{\varepsilon}_i^\mathrm{p}\rVert  -r_i,
	\label{eq103}
\end{align}
\begin{align}
	\phi_{ij} = \dfrac{\partial \Psi_{\mathrm{rel}}}{ \partial\lambda_i}-\dfrac{\partial \Psi_{\mathrm{rel}}}{ \partial\lambda_j}
	- r_i \lVert \boldsymbol{ \varepsilon}_j^\mathrm{p}-\boldsymbol{\varepsilon}_i^\mathrm{p} \rVert.
	\label{eq104}
\end{align}
Whereas the first one is the well-known stress function for plastic evolution, the second one represents the yield condition for transition between the phases. 

\subsection{Initiation of new phase}

Physically, if is not possible to determine an initial value of the plastic strain. Only its evolution is given. This raises the question at which value of plastic strain a new phase will be initiated. We assume here new phases being created as soon as thermodynamically possible. Mathematically, this means that if a phase $j$ is newly formed via a phase transition from phase $i$ to phase $j$, phase $j$ will be created with plastic strain determined as
\begin{align*}
\boldsymbol{\varepsilon}_{j,\mathrm{ini}}^\mathrm{p} = \arg \max \biggl\{\phi_{ij} \bigg\vert _{\lambda_i=0} \vert \boldsymbol{\varepsilon}_j^\mathrm{p}  \biggl\}.
\end{align*}
Related approaches to phase initiation can be found in \cite{Kochmann1,Hackl2012}.

\section{Numerical results}\label{sec3}
The incremental update procedure employed is shown in Algorithm \eqref{alg:cap}. A similar procedure was already used in \cite{Kochmann1} for modeling the evolution of first order laminates. \\
Our subsequent examples are based on a material with three different phases in an displacement driven setting. 
The material point behaviour is investigated first, followed by a two dimensional Finite Element (FE) benchmark test.

\subsection{Material point behaviour}

\begin{figure}[H]
\centering
   \begin{minipage}[b]{.49\linewidth} % [b] => Ausrichtung an \caption
   \centering
      \includegraphics[width=\linewidth]{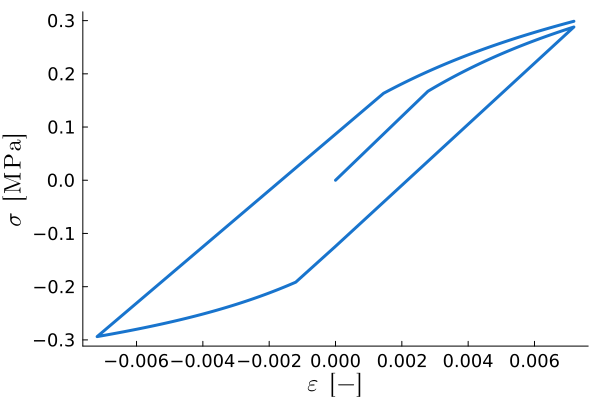}
      (a)
   \end{minipage}
   \hspace{.01\linewidth}% Abstand zwischen Bilder
   \begin{minipage}[b]{.49\linewidth} % [b] => Ausrichtung an \caption
   \centering
      \includegraphics[width=\linewidth]{ 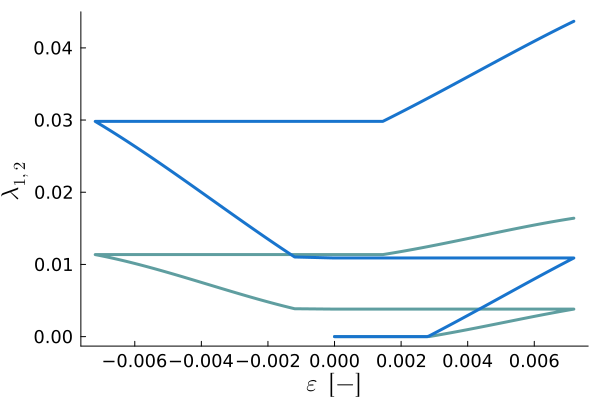}
     (b)
   \end{minipage}
  
 \begin{minipage}[b]{.49\linewidth} % [b] => Ausrichtung an \caption
   \centering
      \includegraphics[width=\linewidth]{ 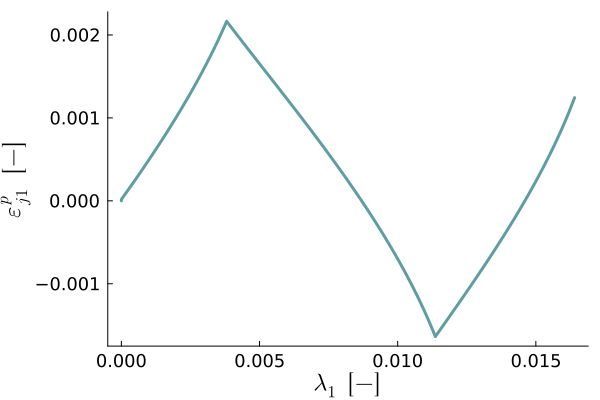}
      (c)
   \end{minipage}
   \hspace{.01\linewidth}% Abstand zwischen Bilder
   \begin{minipage}[b]{.49\linewidth} % [b] => Ausrichtung an \caption
   \centering
      \includegraphics[width=\linewidth]{ 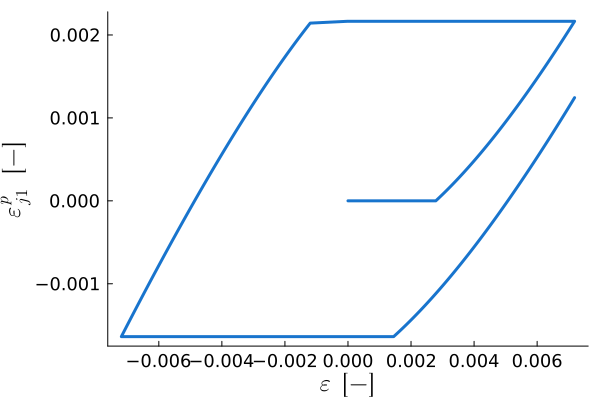}
     (d)
   \end{minipage}

   \caption{\raggedright Material point behaviour during a cyclic tension test. The relationship between stress and strain is compared with the phase development. (a) stress/strain diagramm, (b) phase evolution of new phases $\lambda_1, \lambda_2$, (c) and (d), development of plastic strain of new phase $\lambda_1$. Associated material parameters can be seen in Tab. \eqref{tab1}.}
      \label{fig:1}
\end{figure}

In the following one-dimensional example we restrict ourselves to $k$ = 3. We assume the initial phase fractions as $\lambda_1=\lambda_2=0$ and $\lambda_3$ = 1. The  strain $\varepsilon$ is cycled between $\pm$ 6.5 \%  during tension with a timestep of $\Delta \mathrm{t}$= 0.00046. The viscosity parameters are set to be, $\eta_1$ =1,16$ \cdot 10^{-10}$ [1/Pa] and $\eta_2$ = 6,91$ \cdot 10^{-9}$ [1/Pa], see Eq.~\eqref{eq:eq19}.\\
The associated material parameters are listed in Tab. \eqref{tab1} and the results are depicted in Fig. \eqref{fig:1}.\\
During the test, the phases $\lambda_{1}$ and $\lambda_{2}$ increase immidiately after reaching the yield stress. The gradient and the amount of increase depend on the selected parameters. An interruption in development only occurs in the elastic area and the existing phase developed further thereafter.
\begin{table}[h]
\caption{material point parameters for $i,j \in$(1,..3) }\label{tab1}%
\begin{tabular}{@{}l | c | c | c | c  | c   @{}}
\toprule

&$C_i$ [MPa] & $c_i$ [kPa]& $r_i$ [Pa]& $b_i$ [Pa]&$\lambda_i$ [-]\\
\hline

$\lambda_1$& 21.60 & 0.00  & 0.002& 0.02&0 \\
$\lambda_2$& 40.00 & 0.00  & 0.003 & 0.03& 1 \\
$\lambda_3$& 07.43 & 1.50  & 0.099 & 0.02& 0 \\
\hline

& $C_j$ [MPa] & $c_j$ [kPa]&$r_j$ [Pa]& $b_j$ [Pa]&$\lambda_j$ [-]\\
\hline

$\lambda_1$& 07.43 & -0.81  & 0.005 & 0.04  &0\\
$\lambda_2$&  21.60 &  -7.35  &  0.002 & 0.07&0  \\
$\lambda_3$&  60.00 &  1.50  &  0.050 &  0.05& 1 \\
\botrule
\end{tabular}
\end{table}

The same characteristics can be seen in Fig. \eqref{fig:1} (c), (d). Furthermore a shake down behaviour regarding the phase development was observed,  phase fractions reaching a constant value and a purely elastic behaviour remaining. 

\begin{algorithm}
\caption{Incremental subroutine}\label{alg:cap}
\vspace{0.2cm}
For each timestep: $\mathrm{t_{n+1}= \mathrm{t_n} + \Delta\mathrm{t}}$\\
\hspace{0.1cm}

\hspace*{\algorithmicindent} \textbf{Input:} $\boldsymbol{\varepsilon}_\mathrm{n},~\boldsymbol{\varepsilon}_{i,\mathrm{n}}^\mathrm{p},~ b_{i,\mathrm{n}},~ \lambda_{i,\mathrm{n}},~\boldsymbol{\varepsilon}_{j,\mathrm{n}}^\mathrm{p},~ b_{j,\mathrm{n}},~ \lambda_{j,\mathrm{n}}  $\\
    \hspace{0.1cm}
\begin{algorithmic}
\For{each phase $\lambda_i$}

\If {$\phi_{ij} \geq 0$ }\Comment{\textit{plastic region}}
\vspace{0.2 cm}
\If{$\lambda_i$ = 0}\Comment{\textit{ Initiation :~only one phase exists}}
\vspace{0.2 cm}
\State Calculate: \hspace{0.2 cm}$\dot{\boldsymbol{ \varepsilon}}_{j,\mathrm{ini}}^\mathrm{p}~\rightarrow~\boldsymbol{\varepsilon}_{j,\mathrm{ini}}^\mathrm{p} $ \vspace{0.1cm}
\State \hspace{2.1cm}$~\curvearrowright ~\dot \lambda_j  = \sum\limits_i  g_{ij} \rightarrow \lambda_{j,\mathrm{n}+1} $
\State Calculate: \hspace{0.2 cm} $ \mathbb{C}_{\mathrm{eff, ini}}, ~ \boldsymbol{\varepsilon}_{\mathrm{eff,ini}}^\mathrm{p}$
\Else\Comment{\textit{ more then one phase}}
\State Calculate:\hspace{0.2 cm}$ \boldsymbol{\varepsilon}_{j,\mathrm{n}}^\mathrm{p}$ \hspace{0.1 cm}$\curvearrowright ~\dot \lambda_j = \sum\limits_i  g_{ij}\hspace{0.2cm} \rightarrow \lambda_{j, \mathrm{n}+1} $
\State Calculate: \hspace{0.2 cm} $ \mathbb{C}_{\mathrm{eff, \mathrm{n}+1}}, ~ \boldsymbol{\varepsilon}_{\mathrm{eff, \mathrm{n}+1}}^\mathrm{p}$ 
\vspace{0.1cm}
\EndIf
\vspace{0.2 cm}
\State Calculate:\hspace{0.2cm}$\dot{\boldsymbol{ \varepsilon}}_{j, \mathrm{n}+1}^\mathrm{p} ~\rightarrow~\boldsymbol{ \varepsilon}_{j,\mathrm{n}+1}^\mathrm{p} $
\vspace{0.1cm}
\State Calculate: \hspace{0.2cm}$ \boldsymbol{\sigma}_{\mathrm{n}+1},~ b_{j,\mathrm{n+1}}$
\vspace{0.2 cm}
\State UPDATE 
\State \Return 
\vspace{0.3cm}
\ElsIf{}\Comment{\textit{elastic region}}
\vspace{0.1cm}
 \State Calculate:\hspace{0.2cm}$\boldsymbol{\sigma}_{\mathrm{n}+1},~ \mathbb{C}_{\mathrm{eff, n+1}}, ~ \boldsymbol{\varepsilon}_{\mathrm{eff, n+1}}^\mathrm{p}$
\vspace{0.1cm}
\State UPDATE
\State \Return
\EndIf
\EndFor
\end{algorithmic}
\end{algorithm}

\subsection{Two-dimenional benchmark test}

\begin{figure}[H]
\centering
   \begin{minipage}[b]{.5\linewidth} % [b] => Ausrichtung an \caption
   \centering
      \includegraphics[width=\linewidth]{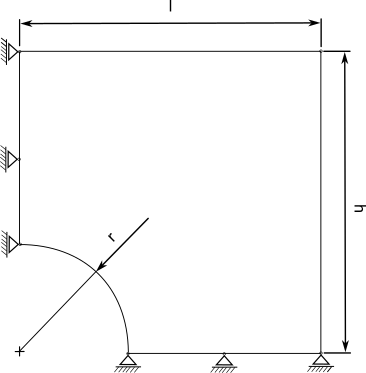}
      
   \end{minipage}
  
   \caption{\raggedright Geometry and boundary conditions of quarter of square plate with a circular hole. }
      \label{fig:2}
\end{figure}

\begin{figure}[H]
\centering
   \begin{minipage}[b]{.8\linewidth} % [b] => Ausrichtung an \caption
   \centering
      \includegraphics[width=\linewidth]{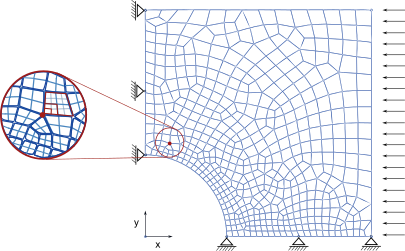}
      
   \end{minipage}
  
   \caption{\raggedright Illustration of three different meshes used for the calculation. A zoom in of the employed mesh with quadrilateral elements shows two stages of refinement. A node marked as a reference point is highlighted in red. }
      \label{fig:3}
\end{figure}

A well-known benchmark boundary value problem from finite element (FE)-calculation is carried out in two dimensions. As can be seen in Fig. \eqref{fig:2} and Fig. \eqref{fig:3} a square plate with a circular hole is subjected to a compressive displacement in x-direction. Due to symmery, it is sufficient to model one quater of the structure. The height and width are assumed to be h = 2.5 m and l = 2.5 m, whereas the radius is r = 0.9 m. The viscosity parameters are $\eta_1$ = 1.60 $ \cdot 10^{-6}$ [1/Pa] and $\eta_2$ = 1.00 $ \cdot 10^{-5}$ [1/Pa]. For the FE-simulation, an unstructured mesh with four-node elements was created. The analyses presented in this paper are carried out with the open source software Gmsh and Julia Ferrite.  \\

\begin{figure}[H]
\centering
   \begin{minipage}[b]{1\linewidth} % [b] => Ausrichtung an \caption
   \centering
      \includegraphics[width=\linewidth]{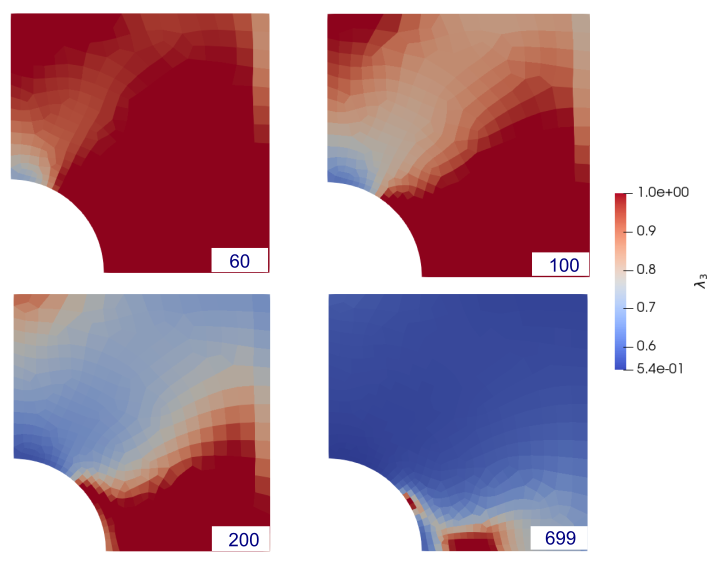}
   \end{minipage}
  
   \caption{\raggedright Development of the original phase during loading at four selected time steps indicated. It can be seen that the phase fraction converges to $\lambda_3\approx 0.54$.}
      \label{fig:4}
\end{figure}

The boundary conditions due to symmtery are
\begin{align*}
u_x &= 0,~ x = 0 \hspace{0.1cm} \mathrm{left side}\\ u_y &= 0,~ y = 0 \hspace{0.1cm} \mathrm{ bottom}.
\end{align*}
The mesh shows a mesh refinement close to the hole and at the bottom to ensure better data evaluation.
All data used for the FE calculations can be seen in Tab. \eqref{tab2}. 
\begin{table}[h]
\caption{Two-dimensional FE-parameters for $i,j \in$(1,..3) }\label{tab2}%
\begin{tabular}{@{}l | c | c | c | c  | c   @{}}
\toprule

&$C_i$ [Pa] & $c_i$ [Pa]& $r_i$ [Pa]& $b_i$ [Pa]&$\lambda_i$ [-]\\
\hline

$\lambda_1$& 50.00 & 28.60  & 0.004& 0.002&0 \\
$\lambda_2$& 30.00 & 19.50  & 0.009 & 0.003& 1 \\
$\lambda_3$& 40.00 & 01.01  & 0.010 & 0.001& 0 \\
\hline

& $C_j$ [Pa] & $c_j$ [Pa]&$r_j$ [Pa]& $b_j$ [Pa]&$\lambda_j$ [-]\\
\hline

$\lambda_1$& 10.00 & -30.30 & 0.003 & 0.003  &0\\
$\lambda_2$&  20.00 &  -02.05   &  0.008 & 0.004&0  \\
$\lambda_3$&  80.00 &  01.08  &  0.002 &  0.002& 1 \\
\botrule
\end{tabular}
\end{table}
To provide a global overview of the phase  development a contour plot of the inital phase $\lambda_3$ is shown in Fig. \eqref{fig:4} depicting four consecutive loading steps. Phase transformation starts at the hole boundary and develops throughtout the sample. The inserted numbers indicate the time steps. For this example, a total of 700 steps are calculated. It can be seen, that finally about half the volume is transformed at $\lambda_3 \approx 0.54$. \\

\begin{figure}[H]
\centering
   \begin{minipage}[b]{.49\linewidth} % [b] => Ausrichtung an \caption
   \centering
      \includegraphics[width=\linewidth]{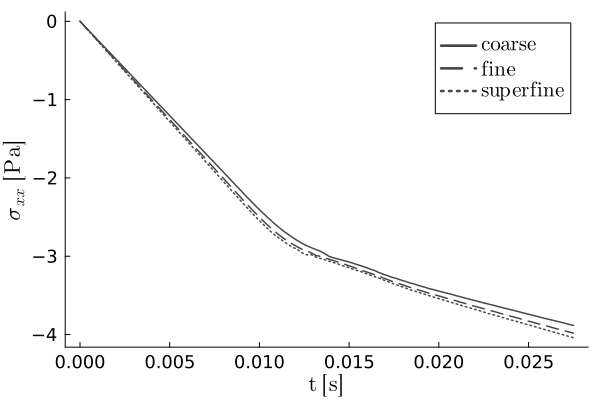} 
      (a)
   \end{minipage}
   \hspace{.01\linewidth}% Abstan zwischen Bilder
    \vspace{0.6cm}
   \begin{minipage}[b]{.49\linewidth} % [b] => Ausrichtung an \caption
   \centering
      \includegraphics[width=\linewidth]{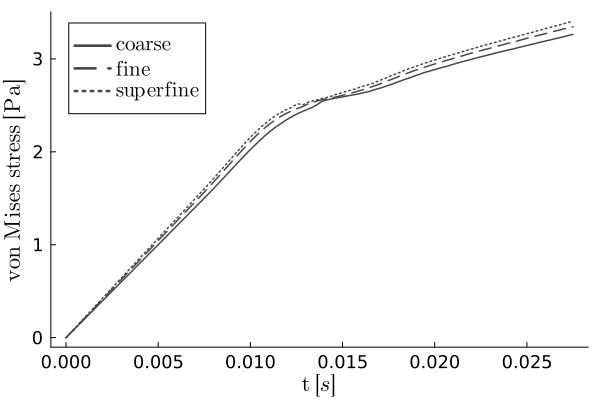}
     (b)
   \end{minipage}
    \begin{minipage}[b]{.49\linewidth} % [b] => Ausrichtung an \caption
   \centering
      \includegraphics[width=\linewidth]{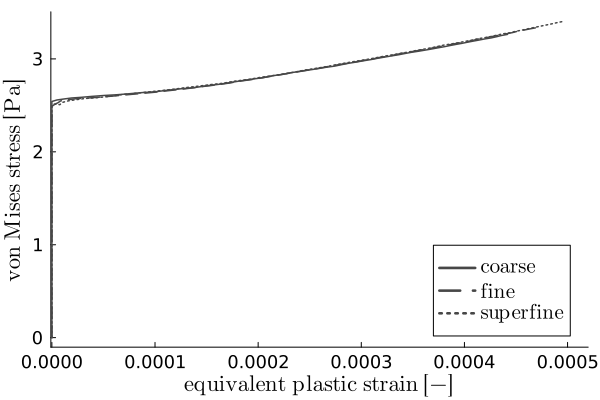}
   (c)
   \end{minipage}
   \hspace{.01\linewidth}% Abstand zwischen Bilder
   \begin{minipage}[b]{.49\linewidth} % [b] => Ausrichtung an \caption
   \centering
      \includegraphics[width=\linewidth]{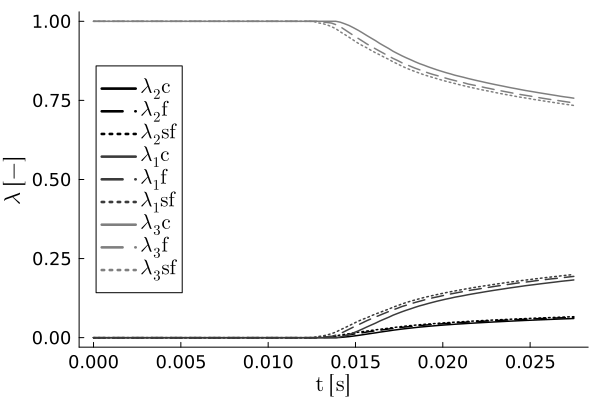}
     (d)
   \end{minipage}

   \caption{\raggedright Comparison of the time development of various quantities at a selected node for three different mesh sizes; coarse (solid), fine (dashed) and superfine (dotted). (a) Normal stress component. (b), (c) equivalent von Mises stress. (d) phase evolution. }
      \label{fig:5}
\end{figure}

A total of three meshes posessing decreasing mesh width are employed, whereby in Fig. \eqref{fig:3} only the coarsest mesh is shown. The zoom-in visualizes the meshes used in this work, which are referred to below as coarse (dark blue), fine (light blue) and superfine (greyish). 
For a comparison, three elements of these meshes are also marked in red, possessing a common node. \\ 
An average value of the element data and thus of the four underlying Gaussian points can be related to the selected node. These averages are depicted in Fig. \eqref{fig:5}. The coarse mesh (solid line), fine mesh (dashed line) and superfine (dotted line) are shown for comparison. \\  
The stress-strain behaviour and the van Mises stress, indicating elastic and plastic behavior, are shown in Fig. \eqref{fig:5} (a)-(c). During the temporal development we observe convergence this can also be seen in the phase development over time in Fig. \eqref{fig:5} (d).
\backmatter

\subsection{Conclusion and outlook}

A model for phase evolution in an elasto-plastic material based on a vritional framework was presented. 
The phase evolution in the microstructure proceeds immediately after entering the plastic region and is only stopped by stress release. The simulations also show a shakedown behaviour during loading, so that the phase generation ends at a certain value.\\ 
This model shows that it is possible to control the phase development via an underlying transition matrix.\\
Hence, with respect to other materials with microstructural properties, additional internal material features have to be added to obtain a realistic physical representation. \\
That is why, in future work in this area, the transition matrix could be extended. For this purpose, as for example in \cite{MWolff,Gov}, a dependence of the transition rate as an exponential function can give access to material internal interactions. 
Moreover, the formulation presented here is related  to isothermal situations. Due to the general approach, it is possible to include laminate structures or other forms of microstructure into this model.
%If your article has accompanying supplementary file/s please state so here. 
%Authors reporting data from electrophoretic gels and blots should supply the full unprocessed scans for key as part of their Supplementary information. This may be requested by the editorial team/s if it is missing.

%Please refer to Journal-level guidance for any specific requirements.

\bmhead{Acknowledgements}
This work has been founded by the Deutsche Forschungsgesellschaft (DFG). We gratefully acknowledge the support within the projekt SPP 2256 ("Variational Methods for Predicting Complex Phenomena in Engineering Structures and Materials"), projekt ID 442307685. \\
Also we thank the Julia community.




%\end{appendices}




\end{document}